# Dutch Disease and the Resource Curse:
# The Progression of Views from Exchange Rates to
# Women's Agency and Well-Being

*Nidhiya Menon, Brandeis University, nmenon@brandeis.edu*

*Yana van der Meulen Rodgers, Rutgers University, yana.rodgers@rutgers.edu*

**Abstract:** This article provides an overview of the history of economic thought on natural resource extraction, which has long been considered an enclave industry with few benefits for areas beyond the local economy. We focus on more recent scholarship examining the social impacts of natural resource extraction, emphasizing gender-related outcomes and determinants. An important lesson from this scholarship is that it is difficult to discuss sustainable development in its contemporary sense without paying due diligence to the gender dimensions of natural resource extraction. A lesson highlighted is that the "resource curse" view of natural capital may not be as pervasive as previously thought.

# Introduction

Natural resource extraction has long been considered an enclave industry which benefits the local economy at the expense of other regions within a country. Combined with political economy considerations in which investment in the extraction industry is prioritized over other sectors, these adverse effects have contributed to the view of natural resource extraction as a disease and a curse (Bebbington et al. 2008). More specifically, the term "resource curse" was introduced in Auty (1993) to describe the lack of significant or prolonged economic growth in many resource-rich developing nations. Typically, the resource curse refers to a paradox in which countries well-endowed with natural resources tend to experience slower economic growth compared to countries with fewer natural resources.

Research on this topic started after the Netherlands enjoyed an economic boom following natural gas discoveries in the North Sea in the 1950s and 1960s. However, foreign currency inflows from the windfall caused higher inflation, real exchange rate appreciation, and slower growth for other sectors, especially manufacturing. Subsequent scholarship on natural resource extraction broadened the scope of analysis to offer interesting correlations on outcomes such as political corruption, environmental degradation, social disruption, political instability, economic instability and conflict (Ross 2012; Andersen and Ross 2014; Ross 2015; Berman et al. 2017). Much of this research framed extraction and mining as a resource curse.

However, recent studies that use rigorous methods to facilitate causal interpretations have challenged this view. The focus of these studies has changed from macro-level variables to micro-level outcomes such as household expenditures and child mortality, with findings that natural resource extraction can indeed serve as a pathway to improved social well-being. These newer studies have gone beyond correlational analysis to evaluate a number of individual and gender-related outcomes including women's employment, women's decision-making power in and outside the home, and gender-based violence (e.g. Kotsadam and Benshaul-Tolonen 2016; Guimbeau et al. 2023).

In the remainder of this article, we offer some highlights of the history of economic thought on Dutch disease and the resource curse. We also review the more recent studies on social outcomes and address the conditions under which natural resource extraction can be consistent with sustainable development goals. We argue that the jury is still out on whether the extraction industry empowers or disempowers vulnerable groups such as women, especially when it comes to intra-household dynamics and tradeoffs between wealth and health. Although some evidence highlights the benefits for women of resource extraction, especially when their labor is valued, these results must be placed in the specific context of socio-cultural norms and attitudes about women's roles in which the studies were conducted, which, in the end, may determine whether any gains made are long-lasting.

In the next section, we begin by describing studies that involve national macroeconomic and governance dynamics. The section that follows describes studies that adopt a more microeconomic lens to document beneficial but localized spillovers from natural resource extraction. Circling back to the discourse on development, this discussion highlights that depending on the lens we use, the "resource curse" may not be as prevalent and widespread as previously believed.

## From Dutch Disease to the Resource Curse

The term 'Dutch disease' was coined by *The Economist* magazine in 1977 when it wrote about the economic crisis in the Netherlands following the discovery of large natural gas deposits in the North Sea in 1959. The newfound wealth from natural gas exports led to a sharp appreciation of the Dutch Guilder, making other Dutch exports less competitive and causing a decline in the manufacturing sector. Subsequently, the UK, Australia, and many developing countries experienced similar problems with their natural resource and mining booms (Mien and Goujon 2021).

In the development economics literature, Dutch disease came to refer more broadly to situations where a sudden resource boom (such as oil, gas, or mineral discoveries) led to economic challenges, especially when it made an economy overly reliant on one sector at the expense of other critical sectors. Corden and Neary (1982) developed a rigorous classic economic model describing Dutch disease, focusing on the effects of a booming resource sector on the rest of the economy. In their model, the Disease comprised two main effects. First, in a resource movement effect, labor shifted from the lagging sector (usually manufacturing) to the booming sector (natural resources), causing direct deindustrialization. Second, in a spending effect, increased revenue from the resource boom raised demand for non-tradable goods, leading to indirect deindustrialization and an increase in the real exchange rate. This study laid the groundwork for understanding the economic dynamics of resource booms and their broader implications. Neary and van Wijnbergen (1986) extended this theoretical framework of Dutch disease, analyzing the role of government policy in mitigating its adverse effects.

The resource curse phenomenon is often attributed to exchange rate appreciation, which focuses investments in a single sector of the economy. It is also linked to subsequent poor governance, corruption, and conflict, which can arise from the mismanagement of resource wealth. For example, the mining industry in India has been causally linked to the election of criminal politicians when the value of minerals in their area increases unexpectedly due to local resource rent shocks (Asher and Novosad 2023). There is a further compounded effect as such politicians commit additional crimes to aggregate wealth when the value of minerals continues to rise during their tenure in office (Asher and Novosad 2023). Although the concept of the resource curse is more recent than Dutch disease, it has since encompassed it, with Dutch disease now often considered one of the mechanisms through which the resource curse operates. The development economics literature now has an abundance of rigorous studies on Dutch disease and the resource curse, as well as several informative and thorough review articles on this scholarship (van der Ploeg 2011; Gilberthorpe and Papyrakis 2015; Badeeb et al. 2017; Mien and Goujon 2021).

## Challenges to the Conventional Wisdom

Fairly recent studies from a number of developing countries have challenged the conventional wisdom that natural resources act more as a curse than a blessing, and have found evidence of improvements in living standards and positive spillover effects for local employment. For instance, using a district-specific treatment and control design as well as exogenous news shocks about discoveries, Mamo et al. (2019) reported significant improvements in living standards, as indicated by night-lights, in districts with new large-scale mining operations, though these benefits did not extend to other districts. Fafchamps et al. (2017), using panel data with a fixed-effects framework, found that areas with gold mines exhibited signs of 'proto-urbanization,' with more advanced economic activities and higher population densities. Also related to an improvement in living standards, Wilson (2012), using a difference-in-differences style approach, causally determined that the copper price boom from 2003 to 2008 led to a decline in risky sexual behavior in Zambian copper mining cities, and Benshaul-Tolonen (2018) and Benshaul-Tolonen et al. (2019), exploiting plausibly exogenous spatial-temporal variation, demonstrated that open-pit gold mining reduced child mortality rates.

Also examining spillover effects, Lippert (2014) used exogenous variation in mine-level output to evaluate the resource boom in Zambia and found that an increase in local copper production increased household expenditures and improved other well-being measures for households close to mines. In related work, Aragón and Rud (2013) leveraged heterogenous exposure to an exogenous demand shock to document positive backward linkage from a gold mine to surrounding areas in Peru in terms of increased employment and real income.

This research on the social dimensions of natural resource extraction has extended to analyses of women's employment and agency. In sub-Saharan Africa, geographic difference-in-differences estimates indicate that industrial mining operations have shifted women's employment from agricultural self-employment to wage employment and service-sector jobs associated with the mining industry (Kotsadam and Benshaul-Tolonen 2016). Also in sub-Saharan Africa, data on precise mine location and year of mine opening capture spatial and temporal variation in the spread of industrial-scale gold mines, revealing improvements in women's access to health care, an increase in women's employment in the service sector, and lower tolerance of domestic violence (Benshaul-Tolonen 2024). Job creation for women in the service sector due to gold mining is not unique to sub-Saharan Africa. In the U.S., mediation analysis applied to a century of census data shows that the Gold Rush of the mid-1800s led to a persistent increase in women's participation in service sector jobs (Aguilar-Gomez and Benshaul-Tolonen 2023). And in an empirical study of the impact of mineral mining on women's agency in India that uses a saturated fixed-effects design, Guimbeau et al. (2023) found that women in proximity to mines that employ relatively high shares of their labor (which, in turn, depends on the exogenously-occurring mineral deposit present), exhibit less tolerance of physical violence and report fewer barriers to accessing healthcare. At the same time, men's likelihood of making decisions jointly with such spouses increases, and men are less likely to be accepting of domestic violence. A key mechanism for these beneficial results is profit-sharing with local communities. More specifically, legislation that requires mining companies to re-invest a fixed proportion of their profits back into local communities, where decisions on how those resources are to be spent are made in conjunction with community and NGO leadership, is found to yield the largest dividends to populations exposed to resource development (Guimbeau et al. 2023).

As with many features of economic development, there are tradeoffs, and the recent scholarship on the social dimensions of resource extraction has uncovered some unpalatable adjustments. For example, resource booms have been associated with net income losses in the US (Jacobsen et al. 2023), increased interpersonal conflicts in Congo (Parker and Vadheim 2017), and a loss in agricultural productivity due to environmental pollutants (see Aragón and Rud 2016 for a careful evaluation of Ghana's large-scale gold mining industry). These tradeoffs also extend to women's outcomes, including higher rates of domestic violence against women in sub-Saharan Africa, where physical violence against women is more commonly accepted, as documented in Kotsadam et al. (2017) using georeferenced data on mine openings and closings matched with individual-level data on abuse.

Alternatively, using panel data on the closure of coal mines in the UK matched to census data from 1981-2011 and a difference-in-differences framework, Aragón et al. 2018 find that women in the manufacturing sector experienced employment losses with the collapse of the coal industry. These results suggest that negative effects on women's outcomes from resource booms and busts are not restricted to a single context, but span countries along the development spectrum.

Tradeoffs are also found between wealth and health: using data from 44 developing countries and difference-in-differences tests, von der Goltz and Barwal (2019) find that mines causally contribute to economic growth, but they come with significant health risks. Women living near heavy metal mines face a higher risk of anemia and blood toxicity from heavy metals such as lead, while children are more likely to experience stunted growth. These adverse effects are primarily found in areas close to mines where metal contamination is likely (von der Goltz and Barwal 2019). Guimbeau et al. (2023) also uncover a tradeoff with women's health: women living close to mines in India are less likely to be underweight, but have a greater incidence of mild anemia possibly due to the higher pollution levels near mines. This tradeoff is consistent with the main conclusion of a literature review in Baum and Benshaul-Tolonen (2021) on mining and gender equality: extractive industries have distinct gender-specific impacts, and health outcomes, including sexual, reproductive, and infant health, are shaped by environmental factors like air and water pollution.

The overall argument is that we need a closer examination of the conditioning factors that predict different types of outcomes from resource extraction, and while the curse may continue to prevail, the microeconomics literature indicates that economic opportunities that mineral extraction affords may serve to mitigate the negative effects even if not fully reversing them. Based on the evidence discussed above, we hypothesize that (a mix of) factors such as the identity of winners and losers (whether individuals or firms reap the benefits), whether negative environmental spillovers supersede accrued returns, whether legislation favors communities and ensures equitable sharing of rents, and the cultural context including accepted local social norms, are important determinants of the net benefits of resource extraction.

**Moving Forward: Gender-Aware Scholarship and Policymaking on Resource Extraction**

Examining the conditions under which resource extraction enhances women's empowerment and human capital can reveal the potential advantages of an industry often seen as exploitative and resource-depleting. Fairly recent scholarship in development economics demonstrates that contrary to the conclusions of early studies on Dutch disease and the resource curse, the natural resources industries can indeed play a role in sustainable development and social well-being. In particular, there is now rigorous evidence that the mining sector has the potential to bring inclusive economic opportunities and social change for women in mining communities. Thus, the commonly accepted "resource curse" view of natural capital may not be as pervasive as previously believed. There are, however, adjustments that need further examination with robust empirical strategies to help guide appropriate policy measures.

Such measures include those relevant to formalized corporations such as capacity-building programs to boost women's employment, training, and mentoring to help them attain higher positions within the mining industry, ensuring equal pay for equal work, improving working conditions, and strictly enforcing anti-harassment policies (Eftimie et al. 2009a). Policies to protect vulnerable populations should also involve initiatives with government, mining industry, and civil society stakeholders. An example is gender-sensitive training programs for service providers (Eftimie et al. 2009b). Additionally, enforcing legislation that mandates profit-sharing with community groups can be transformational (Guimbeau et al. 2023).

Although mining remains predominantly male-dominated, women and girls are increasingly participating in artisanal and small-scale mining (Bashwira et al. 2014). Emphasizing community dialogues and participatory planning in both large and small mining projects can amplify the voices of local women workers, thereby ensuring that the industry generates positive socio-economic spillovers (Pokorny et al. 2019). Developing capacity-building programs and fostering participatory engagement are policy solutions relevant for both large-scale extraction and artisanal mining that may be customized to fit local needs and conditions.

Such policies are critical for fostering an environment where women can be equal partners in contributing more effectively to the economy, thereby driving inclusive economic growth which is free from local capture. Strengthening institutions to enforce legislation in mining areas aligns with efforts to ensure that mining benefits the broader economy (Mehlum et al. 2006). There is growing interest among researchers and policymakers to reframe mining from an isolated sector with negative economic impacts to one that engenders societal well-being, and which may be leveraged to further expand women's agency at home and in the workplace.

Bashwira, Marie-Rose, Jeroen Cuvelier, Dorothea Hilhorst, and Gemma Van der Haar. 2014. "Not only a man's world: Women's involvement in artisanal mining in Eastern DRC." *Resources Policy* 40: 109-116.

Baum, Sarah, and Anja Benshaul-Tolonen. 2021. "Extractive industries and gender equality." *Review of Environmental Economics and Policy* 15 (2): 195-215.

Bebbington, Anthony, Leonith Hinojosa, Denise Humphreys Bebbington, Maria Luisa Burneo, and Ximena Warnaars. 2008. "Contention and ambiguity: Mining and the possibilities of development." *Development and Change* 39 (6): 887-914.

Benshaul-Tolonen, Anja. 2024. "Industrial gold mining and female empowerment." *Economic Development and Cultural Change* 72 (3): 1213-1266.

Benshaul-Tolonen, Anja. 2018. "Local industrial shocks and infant mortality." *The Economic Journal* 129 (620): 1561-1592.

Benshaul-Tolonen, Anja, Punam Chuhan-Pole, Andrew Dabalen, Andreas Kotsadam, and Aly Sanoh. 2019. "The local socioeconomic effects of gold mining: Evidence from Ghana." *The Extractive Industries and Society* 6 (4): 1234-1255.

Berman, Nicolas, Mathieu Couttenier, Dominic Rohner, and Mathias Thoenig. 2017. "This mine is mine! How minerals fuel conflicts in Africa." *American Economic Review* 107 (6): 1564-1610.

Corden, W. Max, and J. Peter Neary. 1982. "Booming sector and de-industrialisation in a small open economy." *The Economic Journal* 92(368): 825–848.

Eftimie, Adriana, Katherine Heller, and John Strongman. 2009a. *Mainstreaming Gender into Extractive Industries Projects: Guidance Note for Task Team Leaders*. Report. Washington, DC: World Bank.11